\documentstyle[aps,prl]{revtex}

\input psfig.tex
\begin{document}

\def\slash#1{\setbox0=\hbox{$#1$}#1\hskip-\wd0\hbox to\wd0{\hss\sl/\/\hss}}

\title{Energy transfer in two-dimensional magnetohydrodynamic turbulence}
\author{Gaurav Dar\thanks{e-mail: gdar@iitk.ac.in} and
Mahendra K. Verma  \\ 
Department of Physics \\ 
Indian Institute of Technology, 
Kanpur 208016, India \\
V. Eswaran  \\ 
Department of Mechanical Engineering \\ 
Indian Institute of Technology, 
Kanpur 208016, India}
\maketitle

\begin{abstract}
In an earlier paper we had developed a method
for computing the effective energy transfer between any two
Fourier modes in fluid or magnetohydrodynamic (MHD) flows. 
This method is applied to a pseudo-spectral, direct numerical simulation
(DNS) study of energy
transfer in the quasi-steady state of 2-D MHD 
turbulence with large scale kinetic forcing. Two aspects of energy
transfer are studied: the energy fluxes, and the energy transfer
between different wavenumber regions ({\it shells}). 
The picture of energy fluxes that emerges is quite complex --- there is
a forward cascade of magnetic energy, 
an inverse cascade of kinetic energy, a flux of energy from the kinetic to 
the magnetic field, and a reverse flux which transfers the energy
back to the kinetic from the magnetic. The energy transfer between different 
wave number shells is also complex --- local and nonlocal transfers often
possess opposing features, i.e.,
energy transfer between some wave number shells occurs
from kinetic to magnetic, and between other wave number shells
this transfer is reversed. The net transfer of energy is from 
kinetic to magnetic. The results obtained from the flux studies and
the  shell-to-shell energy transfer studies are consistent with each
other.
\end{abstract}

\bibliographystyle{unsrt}

\newpage

\section{Introduction}
\label{s:intro}

   In magnetohydrodynamic (MHD) turbulence several scales interact amongst 
themselves
and energy is transferred between them. This transfer plays an important
role in the generation of magnetic fields. In this paper we study the 
energy transfer in two-dimensional(2-D) MHD turbulence.  The MHD and  
Navier Stokes 
equations show that energy is transferred, in spectral space, to a mode 
$\bf k$ from modes ${\bf p}$ and ${\bf q}$ such that the three 
wave numbers satisfy the condition ${\bf k+p+q=0}$. In an earlier paper
(we will refer to this paper as Paper I) \cite{Dar2000_1} we had 
introduced the idea of {\it effective} energy transfer between a pair
of modes within a triad by the mediation of the third mode, and found the
formulae for computing these transfers. In this paper we have applied
this method to study energy transfer in 2-D MHD turbulence.

  In fluid turbulence the dynamics of energy transfer has been well studied. 
In 3-D fluid turbulence the kinetic energy is transferred from large 
scales to small scales, whereas in two dimensions there is an inverse 
cascade of kinetic energy from small scales to large scales \cite{Lesieur}. 
For MHD turbulence there have been various phenomenological 
\cite{Kraichnan65,Dobrowolny80,Zhou90b},
analytical \cite{Pouquet78,Ishizawa298,Pouquet76,Grappin182,Grappin83,Camargo92,MKV95b,MKV99}, 
and numerical studies \cite{Fyfe76,Pouquet88,Politano89,Biskamp89,Kinney95,MKV96,Ishizawa198,Politano98,Basu98,Muller99}
to investigate energy spectra, energy cascade rates, etc.
Contrary to fluid 
turbulence,
the direct numerical simulations (DNS) of MHD turbulence show that the 
{\it total} energy is transferred from large scales to small scales 
both in 2-D as well as 3-D
turbulence \cite{MKV96,MKVThesis}.  There have been theoretical
predictions of the magnitude and directions of only a few of the various 
fluxes  \cite{Pouquet78,MKV95b,Pouquet76}.

 Magnetohydrodynamic turbulence is sometimes 
described 
in terms of the Elsasser variables ${\bf z^{\pm} = u \pm b}$. It has been a 
basic assumption of the phenomenologies for MHD turbulence that the energies 
associated with ${\bf z^{\pm}}$ are transferred to the small scales, and that 
the flux of the energies is constant in the inertial range in both
2-D and 3-D turbulence \cite{Zhou90b,Marsch90_coll}.
Thus the the physics of energy cascades appears to be similar in 2-D
and 3-D MHD turbulence.

 The magnetic energy of a  mode evolves due to two nonlinear 
terms in the
MHD equations---${\bf [b.(u. \nabla) b]}$ and ${\bf [b.(b. \nabla) u]}$---%
the first exchanges magnetic energy between different scales, and the
second exchanges magnetic and kinetic energy between different scales.
The kinetic energy similarly evolves due to two nonlinear terms---%
${\bf [u.(u. \nabla) u]}$ and ${\bf [u.(b. \nabla) b]}$---the first one
exchanges kinetic energy between different scales and the second 
exchanges energy between the magnetic and the velocity fields.  
In this paper we have carried out a detailed investigation
of the various energy transfers in 2-D MHD turbulence.

Pouquet {\it et al.} ~\cite{Pouquet76} studied the energy transfer between 
large and small scales using 3-D EDQNM closure 
calculations. They found that the
large scales of the magnetic field gain energy in 
the presence of small-scale residual helicity (which is the difference 
between kinetic helicity and magnetic helicity).
The large scale magnetic field, in turn, was
found to enhance the energy exchange between the small-scale magnetic 
and the velocity fields. This energy exchange resulted
in equipartition of small-scale kinetic and magnetic energies. 
By drawing an analogy between the MHD equation 
for the magnetic field and the vorticity equation for fluids, 
Batchelor ~\cite{Batchelor50} argued that the transfer between kinetic 
and magnetic energies takes place primarily at small scales. 
A similar conjecture was made by Pouquet and Patterson ~\cite{PouquetPatt78}.
They also conjectured that an inverse cascade of energy from small-scale
to large-scale magnetic field is the mechanism responsible for the
ehancement of large-scale magnetic energy.
Contrary to the prediction from 3-D EDQNM of Pouquet {\it et al.}
 ~\cite{Pouquet76},
simulations of decaying turbulence by Pouquet and Patterson 
 ~\cite{PouquetPatt78} showed that it is the
magnetic helicity and {\it not} the residual helicity which is
important for the growth of large-scale magnetic field.

In a 2-D EDQNM study, Pouquet ~\cite{Pouquet78} obtained 
eddy viscosities for MHD turbulence. She found that
the small-scale magnetic energy acts like a negative eddy viscosity
on the large-scale magnetic energy. The inverse cascade of the mean-square
magnetic vector potential,
and hence the enhancement of large-scale magnetic energy, was
conjectured to arise due to the destablization of the
of large-scale magnetic field by the small-scale magnetic field. She
also found that the small-scale kinetic energy has the effect of a positive
eddy viscosity on the large-scale magnetic energy.
Ishizawa and Hattori ~\cite{Ishizawa298} in their EDQNM calculation
obtained the eddy viscosity due to each of the nonlinear terms in 
MHD equations.
They found that the eddy viscosity due to ${\bf (b.\nabla)u}$ is
positive, leading to a transfer of energy from large-scale magnetic field
to small-scale magnetic field. In their calculation, energy was also found
to be transferred from the small-scale velocity field to the 
large-scale magnetic field. 
Both, Pouquet's ~\cite{Pouquet78} and Ishizawa and Hattori's 
 ~\cite{Ishizawa298}  calculations give a non-local
energy transfer from small-scale velocity field to the large-scale
velocity field.

In a recent work Ishizawa and Hattori
 ~\cite{Ishizawa198} employed the wavelet basis to 
investigate energy transfer in 2-D MHD turbulence. There are some
similarities between the observations made by us in this paper 
and those in the work of Ishizawa and Hattori.
Using the formalism developed in Paper I
we do a thorough investigation of all the energy transfers.
The wavelet basis used by
Ishizawa and Hattori ~\cite{Ishizawa198} is a useful tool for investigating
spectral properties and energy transfer properties in different
regions of the flow.  In Ishizawa's work, the flow was divided into a
turbulent and a coherent region which are distinguished by the level of 
vorticity fluctuations in that region --- turbulence is known to have high
levels of vorticity fluctuations ~\cite{Lumley}. They found that the
energy transfer between velocity/magnetic scales occurs more 
efficiently in the turbulent region and less efficiently in the 
coherent region.

In some of the above mentioned studies \cite{Pouquet76,PouquetPatt78} a 
distinction was not made between
the energies transferred to a mode from the velocity and the
magnetic fields 
Also, the energy transferred into a mode {\bf k} from different wave number 
regions was not separately computed---only the {\it net} 
energy transfer into a wave number {\bf k} was obtained. 
The Eddy damped quasi-normal Markovian (EDQNM) closure
calculations dealt mainly with coarse-grained energy transfer (between
large scales and small scales). In another study, Frick and Sokoloff 
\cite{Frick98} solved
a shell model of MHD turbulence and calculated only the kinetic energy
fluxes between the velocity modes, and the magnetic energy fluxes
between the magnetic modes.
In our simulations we investigate the 
(1) various energy fluxes arising within and between the velocity and
magnetic fields, and (2) the fine-grained (considering many wave number shells)
energy transfer between magnetic and 
kinetic energies. This gives a more informed picture of the physics of 
energy transfer in MHD turbulence.

In 3-D MHD turbulent flows, beyond a critical magnetic
Reynold number, the magnetic field can grow and reach a steady state
\cite{PouquetPatt78,Miller96,Meneguzzi81,Leorat81}. This is called the
{\it dynamo effect}, and is believed to be the mechanism for the generation
of magnetic fields within astrophysical objects.
The non-linear transfer in MHD turbulence is important for 
understanding the growth process of magnetic energy in a dynamo.
Although it is not
possible to indefinitely maintain a steady state magnetic field in 
two-dimensional turbulence \cite{Zeldovich57}, a steady state can be achieved
for a {\it finite} period of time \cite{Pouquet78}. This, coupled with the 
similarity in physics, allows us to probe the non-linear energy transfer
in real MHD flows through 2-D simulations. The methods used here are
completely generalizable to the 3-D case and can be used to 
study the dynamo problem directly at a later date.

The paper is organized into the following sections. In 
Section ~\ref{s:formalism}
we discuss the various energy fluxes and shell-to-shell energy transfer
rates which were obtained in Paper I and which we will compute in this paper.
The numerical methods used are outlined in Section ~\ref{s:sim_details}. 
In subsection ~\ref{subs:num_flux}, we present a numerical technique 
to compute the fluxes and the shell-to-shell 
energy transfer rates. The results are presented in Section ~\ref{s:results}
which is divided into three parts: Subsection ~\ref{subs:steady_state}
outlines the approach used by us to
obtain a steady state; Subsections ~\ref{subs:flux_results} and 
\ref{subs:shell_results} contain the observations for the
energy fluxes  and the shell-to-shell energy transfer rates, respectively.
The discussion follows in Section ~\ref{s:discussion}.

\section{Formalism and Definitions}
\label{s:formalism}
  The MHD equations in real space are written as

\begin{equation}
\frac{\partial {\bf u}}{\partial t} + ({\bf u.\nabla}){\bf u}  = 
 - {\bf \nabla}p + ({\bf b.\nabla}){\bf b} + \nu {\bf \nabla^{2} u} ,
\label{eq:u_mhdeq}
\end{equation}
and
\begin{equation}
\frac{\partial {\bf b}}{\partial t} + ({\bf u.\nabla}){\bf b}   = 
 ({\bf b.\nabla}){\bf u} + \mu {\bf \nabla^{2} b}
\label{eq:b_mhdeq}
\end{equation}
where ${\bf u}$ and ${\bf b}$ are the velocity and 
magnetic fields respectively, and $\nu$ and $\mu$ are the
fluid kinematic viscosity and magnetic diffusivity respectively.
In Fourier space, the kinetic energy and magnetic energy evolution equations
for a Fourier mode are 

\begin{eqnarray}
\frac{\partial E^{u}(\bf k)}{\partial t} + 2 \nu k^{2} E^{u}({\bf k})&  = &
\sum_{{\bf k+p+q}=0} \frac{1}{2}S^{uu}({\bf k|p,q}) + \\  \nonumber
 & & \sum_{{\bf k+p+q}=0} \frac{1}{2}S^{ub}({\bf k|p,q}) 
\label{eq:eu_mhdeq}
\end{eqnarray}

\begin{eqnarray}
\frac{\partial E^{b}(\bf k)}{\partial t} + 2 \mu k^{2} E^{b}({\bf k}) & = &
\sum_{{\bf k+p+q}=0} \frac{1}{2}S^{bb}({\bf k|p,q}) + \\  \nonumber
 & & \sum_{{\bf k+p+q}=0} \frac{1}{2}S^{bu}({\bf k|p,q}) 
\label{eq:eb_mhdeq}
\end{eqnarray}
where $E^{u}({\bf k})={\bf |u(k)|}^{2}/2$ is the kinetic energy,
and $E^{b}({\bf k})={\bf |b(k)|}^{2}/2$ is the magnetic energy.
The four nonlinear terms  $S^{uu}({\bf k|p,q})$, $S^{ub}({\bf k|p,q})$
$S^{bb}({\bf k|p,q})$ and $S^{bu}({\bf k|p,q})$ are 

\begin{eqnarray}
      S^{uu}({\bf k|p,q}) & = 
                         -  Re ( & i{\bf \left[k.u(q)\right]} 
                             {\bf \left [u(k).u(p)\right]} +    \\
                           &     &   i{\bf \left [k.u(p)\right]} 
                                {\bf \left [u(k).u(q)\right]} ) \nonumber
\label{eq:ukpq_mhd_def}
\end{eqnarray}

\begin{eqnarray}
      S^{bb}({\bf k|p,q}) & = 
                        -  Re ( & i{\bf \left [k.u(q)\right]} 
                               {\bf \left [b(k).b(p)\right ]} +  \\  
                         &  &          i{\bf \left [k.u(p)\right]} 
                                    {\bf \left [b(k).b(q)\right]} ) \nonumber
\label{eq:bkpq_mhd_def}
\end{eqnarray}
\begin{eqnarray}
      S^{ub}({\bf k|p,q})  & = 
                          Re ( &  i{\bf \left [k.b(q)\right]}  
                              {\bf \left [u(k).b(p)\right ]} +  \\  
                            &  &       i{\bf \left [k.b(p)\right]} 
                                    {\bf \left [u(k).b(q)\right]} ) \nonumber
\label{eq:ubkpq_mhd_def}
\end{eqnarray}

\begin{eqnarray}
      S^{bu}({\bf k|p,q}) & =  
                          Re ( & i{\bf \left [k.b(q)\right]} 
                               {\bf \left [b(k).u(p)\right ]} +  \\  \nonumber
                         &  &          i{\bf \left [k.b(p)\right]} 
                                    {\bf \left [b(k).u(q)\right]} )
\label{eq:bukpq_mhd_def}
\end{eqnarray}
These terms are  conventionally taken to represent the nonlinear
transfer from modes ${\bf p}$ and ${\bf q}$ to mode ${\bf k}$ 
\cite{Lesieur,Stanisic}.
Note that the wavenumber triad $\bf k$, $\bf p$ and $\bf q$ should satisfy
the condition ${\bf k+p+q}=0$. 
The term $S^{uu}(\bf k|p,q)$ represents the net transfer of kinetic energy
from modes ${\bf p}$ and ${\bf q}$ to mode ${\bf k}$. Likewise the term
$S^{ub}({\bf k|p,q})$ is the net magnetic energy transferred from modes
${\bf p}$ and ${\bf q}$ to the kinetic energy in mode ${\bf k}$,
whereas $S^{bu}({\bf k|p,q})$ is the net kinetic energy transferred from
modes ${\bf p}$ and ${\bf q}$ to the magnetic energy in mode ${\bf k}$.
The term $S^{bb}({\bf k|p,q})$ represents the transfer of magnetic 
energy from modes ${\bf p}$ and ${\bf q}$ to mode ${\bf k}$.  
Thus, the quantities $S^{uu}({\bf k|p,q})$, $S^{ub}({\bf k|p,q})$, 
$S^{bb}({\bf k|p,q})$, and $S^{bu}({\bf k|p,q})$ represent the 
nonlinear energy transfer from the {\it two modes} $\bf p$ and $\bf q$ 
to mode $\bf k$. We define the following quantities :
\begin{equation}
{\cal{\slash{S}}}^{uu}({\bf k|p|q}) = - Re \left( i{\bf \left[k.u(q)\right]} %
{\bf \left[u(k).u(p)\right]} \right) ,
\label{eq:Sukup_def}
\end{equation}    
\begin{equation}
{\cal{\slash{S}}}^{bb}({\bf k|p|q}) = - Re \left( i{\bf \left[k.u(q)\right]} %
{\bf \left[b(k).b(p)\right]} \right) ,
\label{eq:Sbkbp_def}
\end{equation}
\begin{equation}
{\cal{\slash{S}}}^{ub}({\bf k|p|q}) =  Re \left( i{\bf \left[k.b(q)\right]} %
{\bf \left[u(k).b(p)\right]} \right) ,
\label{eq:Sukbp_def}
\end{equation}
\begin{equation}
{\cal{\slash{S}}}^{bu}({\bf k|p|q}) =  Re \left( i{\bf \left[k.b(q)\right]} %
{\bf \left[b(k).u(p)\right]} \right) .
\label{eq:Sbkup_def}
\end{equation}.
where Eqs.~(\ref{eq:Sukup_def})-(\ref{eq:Sbkup_def})
 are the first terms on the right hand side of 
Eqs. (5)-(8).  In Paper I, we showed that the the expressions in 
Eqs.~(\ref{eq:Sukup_def})-(\ref{eq:Sbkup_def}) have the following
intepretation : ${\cal{\slash{S}}}^{uu}({\bf k|p|q})$ is the kinetic energy
{\it effectively} transferred from mode ${\bf p}$ to mode ${\bf k}$; 
${\cal{\slash{S}}}^{bb}({\bf k|p|q})$ is the magnetic energy 
effectively transferred from mode ${\bf p}$ to mode ${\bf k}$; 
${\cal{\slash{S}}}^{ub}({\bf k|p|q})$ is the magnetic energy 
effectively transferred 
from mode ${\bf p}$ to kinetic energy in mode ${\bf k}$, and
${\cal{\slash{S}}}^{bu}({\bf k|p|q})$ is the kinetic energy 
effectively transferred
from mode ${\bf p}$ to magnetic energy in mode ${\bf k}$. In all 
these effective energy transfers the
mode with wavenumber ${\bf q}$ {\it mediates} the transfer but does not
itself gain or lose {\it net} energy. We discussed in Paper I that
Eqs.~(\ref{eq:Sukup_def})-(\ref{eq:Sbkup_def})
 give us a method of computing energy exchange between 
any two modes. This method allows us to have a much better understanding of 
cascade rates between scales.

In this paper, we use the term {\it u}-sphere ({\it b}-sphere) to
represent a sphere in  wavenumber space enclosing velocity (magnetic) modes.
Thus the energy associated with a {\it u}-sphere ({\it b}-sphere) is
kinetic (magnetic) energy. In Paper I, the idea of effective mode-to-mode
transfer rate was used to define the following
energy fluxes and shell-to-shell energy 
transfer rates.
The flux $\Pi^{u<}_{b<}(K)$ is 
the kinetic energy lost by a {\it u}-sphere of radius $K$ to the
magnetic energy of the {\it b}-sphere of the same radius,i.e.,
\begin{equation}
\Pi^{u<}_{b<}(K) =  \sum_{|{\bf k}|<K} \sum_{|{\bf p}|<K} 
                      {\cal{\slash{S}}}^{bu}({\bf k|p|q}) ;
\label{eq:flux_uinbin}
\end{equation}
where the summation is restricted such that ${\bf k+p+q=0}$.
The flux $\Pi^{u<}_{b>}(K)$ is the energy lost by a 
{\it u}-sphere to modes outside the corresponding {\it b}-sphere, i.e.,
\begin{equation}
\Pi^{u<}_{b>}(K) =  \sum_{|{\bf k}|>K} \sum_{|{\bf p}|<K} 
                      {\cal{\slash{S}}}^{bu}({\bf k|p|q}) ;
\label{eq:flux_uinbout}
\end{equation}
The flux $\Pi^{b<}_{u>}(K)$ is the energy lost by a {\it b}-sphere
to modes outside the {\it u}-sphere, i.e.,
\begin{equation}
\Pi^{b<}_{u>}(K) =  \sum_{|{\bf k}|>K} \sum_{|{\bf p}|<K} 
                     {\cal{\slash{S}}}^{ub}({\bf k|p|q}) .
\label{eq:flux_binuout}
\end{equation}
The flux $\Pi^{u>}_{b>}(K)$ is the energy lost by modes outside
a {\it u}-sphere to those outside the corresponding {\it b}-sphere, i.e.,
\begin{equation}
\Pi^{u>}_{b>}(K) =  \sum_{|{\bf k}|>K} \sum_{|{\bf p}|>K} 
                     {\cal{\slash{S}}}^{bu}({\bf k|p|q}) .
\label{eq:flux_uoutbout}
\end{equation}
These are the only fluxes from {\it u} to {\it b} modes. The fluxes
for the kinetic-to-kinetic and magnetic-to-magnetic 
energy transfers can also be written in terms of the mode-to-mode
transfers ${\cal{\slash{S}}}^{uu}$
and ${\cal{\slash{S}}}^{bb}$ respectively.
Flux $\Pi^{u<}_{u>}(K)$ is 
the energy lost by {\it u}-sphere to modes outside the {\it u}-sphere, i.e.,
\begin{equation}
\Pi^{u<}_{u>}(K) =  \sum_{|{\bf k}|>K} \sum_{|{\bf p}|<K} 
                      {\cal{\slash{S}}}^{uu}({\bf k|p|q}) .
\label{eq:flux_uinuout}
\end{equation}
Similarly,  the flux $\Pi^{b<}_{b>}(K)$ is the energy
lost by a {\it b}-sphere to modes outside the {\it b}-sphere, i.e.,
\begin{equation}
\Pi^{b<}_{b>}(K) =  \sum_{|{\bf k}|>K} \sum_{|{\bf p}|<K} 
                      {\cal{\slash{S}}}^{bb}({\bf k|p|q}) .
\label{eq:flux_binbout}
\end{equation}
The total flux is defined as the total energy (kinetic+magnetic) lost
by the $K$-sphere to the modes outside, i.e.,

\begin{equation}
\Pi_{tot}(K) = \Pi^{u<}_{u>}(K) + \Pi^{b<}_{b>}(K) + \Pi^{u<}_{b>}(K) +
                \Pi^{b<}_{u>}(K)  .
\label{eq:flux_tot}
\end{equation}

Similarly, from Paper I, 
the {\it shell-to-shell} energy transfer rate from the $n^{th}$ 
{\it u}-shell to the $m^{th}$  {\it b}-shell is given by 

\begin{equation}
T^{bu}_{mn} = \sum_{{\bf k} {\large \epsilon}{\it m}} \sum_{{\bf p} 
              {\large \epsilon {\it n}}} 
                       {\cal{\slash{S}}}^{bu}({\bf k|p|q}).
\label{eq:bshell_ushell}
\end{equation}
The shell-to-shell energy transfer rate  from the $n^{th}$ {\it u}-shell 
to the $m^{th}$ {\it u}-shell is given by 
\begin{equation}
T^{uu}_{mn} = \sum_{{\bf k} {\large \epsilon}{\it m}} \sum_{{\bf p} 
              {\large \epsilon {\it n}}} 
                       {\cal{\slash{S}}}^{uu}({\bf k|p|q}).
\label{eq:ushell_ushell}
\end{equation}
The shell-to-shell energy transfer rate  from the $n^{th}$ {\it b}-shell 
to the $m^{th}$ 
{\it b}-shell is given by 
\begin{equation}
T^{bb}_{mn} = \sum_{{\bf k} {\large \epsilon}{\it m}} \sum_{{\bf p} 
              {\large \epsilon {\it n}}} 
                       {\cal{\slash{S}}}^{bb}({\bf k|p|q}).
\label{eq:bshell_bshell}
\end{equation}
The energy transfer rate from the $m^{th}$ {\it b}-shell to the
$n^{th}$ {\it u}-shell, $T^{ub}_{nm}=-T^{bu}_{mn}$.

In this paper, we have performed a direct numerical simulation of 
MHD turbulence and
using the values of Fourier modes obtained in the simulation, we have
calculated various fluxes [Eqs.~(\ref{eq:flux_uinbin})-(\ref{eq:flux_binbout})]
and energy transfer rates 
[Eqs.~(\ref{eq:bshell_ushell})-(\ref{eq:bshell_bshell})].
The details of the numerical methods used to compute 
these quantities and to solve the MHD equations are presented in the
following section.

\section{Simulation Details}
\label{s:sim_details}
\subsection{Numerical Method}
\label{subs:num_method}

   In our simulations we use  the Els\"{a}sser variable 
${\bf z}^{\pm}={\bf u} \pm {\bf b}$ instead of {\bf u} and {\bf b}.
The MHD equations (1) and (2) written
in terms of ${\bf z}^{+}$ and ${\bf z}^{-}$ are
\begin{eqnarray}
\frac{\partial{\bf z}^{\pm}}{\partial{t}} \mp 
\left({\bf B}_{0}.{\bf \nabla}\right) {\bf z}^{\pm} +
\left ({\bf z}^{\mp}.{\bf \nabla}\right){\bf z}^{\pm} & = & -{\bf \nabla}p +
 \nu_{\pm} \nabla^{2}{\bf z}^{\pm} + 
\nu_{\mp} \nabla^{2}{\bf z}^{\mp} \nonumber \\
 &  & + \left(\nu_{\pm}/k_{eq}^{2}\right)\nabla^{4}{\bf z}^{\pm} 
      + \left(\nu_{\mp}/k_{eq}^{2}\right)\nabla^{4}{\bf z}^{\mp} +{\bf F^{\pm}}
\label{eq:zpm_mhdeq}
\end{eqnarray}
where  $\nu^{\pm}=(\nu \pm \mu)/2$. 
In addition to the viscosity and the magnetic diffusivity in Eqs.(1) and (2), 
the Eq.~(\ref{eq:zpm_mhdeq})
includes hyperviscosity ($\nu^{\pm}/k_{eq}^{2}$) to 
damp out the energy at very high wave numbers.
We choose $\nu = \mu = 5 \times 10^{-6}$ for runs
on a grid of size $512 \times 512 $. The parameter $k_{eq}$
is chosen to be 14 in all the runs. 
The terms ${\bf F^{\pm}}$ are the forcing
functions. The corresponding forcing functions for the $\bf u$ and the
$\bf b$ fields are related to ${\bf F^{\pm}}$, i.e.,
${\bf F}^{u}=({\bf F}^{+}+{\bf F}^{-})/2$ and
${\bf F}^{b}=({\bf F}^{+}-{\bf F}^{-})/2$ respectively. In our simulations,
we  do not force the magnetic field (i.e, ${\bf F}^{b}=0$). Consequently,
we get ${\bf F}^{+}={\bf F}^{-}={\bf F}^{u}={\bf F}$.

We use the pseudo-spectral method \cite{Canuto}
to solve the above equations in
a periodic box of size $2 \pi \times 2 \pi$. In order to remove the
aliasing errors arising in the pseudo-spectral method a square
truncation is performed wherein all the modes with $|k_{x}| \geq N/3$ or
$|k_{y}| \geq N/3$ are set equal to zero. The equations are time
advanced using the second order Adam-Bashforth scheme for the convective
term and the Crank-Nicholson scheme for the viscous terms.
The time step $\Delta t$ used for these runs is $5 \times 10^{-4}$. 
All the quantities are non-dimensionalised using the initial total energy
and length scale of $2 \pi$. For the simulation results shown in this paper
$\sigma_c \approx 0.1$. However, we have carried out simulations upto
$\sigma_c \approx 0.9$ and the results obtained for higher $\sigma_c$'s
were found to be qualitatively similar to those shown in this paper.

At each time step we construct $\bf F$ as an uncorrelated random function
that is  divergence free (i.e., ${\bf \nabla. F}=0$). The $x$-component 
of $\bf F$ is determined at every time step by
\begin{equation}
F_{x} = \left(\frac{k_{y}}{k}\right) \sqrt{\Delta t} {\cal F} e^{i \phi} ,
\end{equation}
where ${\cal F}^{2}$ is equal to the average energy input rate per 
mode, and
phase $\phi$ is a uniformly distributed random variable between $0$ and $2 \pi$.
The $y$-component of the forcing function is obtained by using the
condition of zero divergence, i.e.,
\begin{equation}
F_{y} = - \left(\frac{k_{x}}{k}\right) \sqrt{\Delta t} {\cal F} e^{i \phi} ,
\end{equation}

We implement the forcing over a k-space annulus $4<k<5$. The value of
${\cal F}^{2}$ is determined from the average rate of the total energy
input which is chosen to be equal to 0.1.

\subsection{Numerical computation of fluxes}
\label{subs:num_flux}

To compute the fluxes we employ a method similar to that
used by Domaradzki and Rogallo \cite{Domaradzki90}.  We outline this 
method below using $\Pi^{b<}_{u>}(K)$ as an example. In 
Eq.~(\ref{eq:flux_binuout})  we 
substitute the expression for $S^{ub}({\bf k|p|q})$ 
[Eq.~(\ref{eq:Sukbp_def}) ] :
\begin{equation}
\Pi^{b<}_{u>}(K) =  \sum_{|{\bf k}|>K} \sum_{|{\bf p}|<K} 
                      Re \left( i{\bf \left[k.b(q)\right]} 
                                   {\bf \left[u(k).b(p)\right]} \right) 
\label{eq:numcomp_binuout}
\end{equation}

A straightforward summation over {\bf k} and {\bf p} involves
$O(N^{2})$ operations, where $N$ is the size of the grid,
and would thus involve a prohibitive computational
cost for high $N$ (i.e., high Reynolds number) simulations. 
Instead, the pseudo-spectral
method can be used to compute Eq.~(\ref{eq:numcomp_binuout}) in
$O(NlogN)$ operations. It involves the following procedure.

We define two `truncated' variables ${\bf u}^{>}$ and ${\bf b}^{<}$ 
as follows
\begin{equation}
 {\bf u}^{>}({\bf k}) = \left \{ \begin{array}{ll}
                              0            & \mbox{if $|{\bf k}|<K$} \\  
                          {\bf u}({\bf k}) & \mbox{if $|{\bf k}|>K$} 
                          \end{array}
                         \right.
\label{eq:utrunc_def}
\end{equation}
and
\begin{equation}
{\bf b}^{<}({\bf p}) = \left \{ \begin{array}{ll}
                          {\bf b}({\bf p}) & \mbox{if $|{\bf p}|<K$}  \\
                              0            & \mbox{if $|{\bf p}|>K$} 
                          \end{array}
                         \right. 
\label{eq:btrunc_def}
\end{equation}
The Eq.~(\ref{eq:numcomp_binuout})
 written in terms of ${\bf u}^{>}$ and ${\bf b}^{<}$ reads as
follows
\begin{equation}
\Pi^{b<}_{u>}(K) =  \sum_{\bf k} \sum_{\bf p} 
                       Re \left( i{\bf \left[k.b(k-p)\right]} 
                              {\bf \left[u^{>}(k).b^{<}(p)\right]} \right) .
\label{eq:flux_binuout_trunc1}
\end{equation}
The above equation may be written as
\begin{equation}
\Pi^{b<}_{u>}(K) =   Re \left [ \sum_{\bf k} i k_{j} u_{i}^{>}({\bf k})
                \sum_{\bf p} b_{j}({\bf k-p}) b_{i}^{<}({\bf p}) \right ]
\label{eq:flux_binuout_trunc2}
\end{equation}
The ${\bf p}$ summation in the equation above can be recognized as a
convolution sum. The right hand side of 
Eq.~(\ref{eq:flux_binuout_trunc2})
can be conveniently and
efficiently evaluated by the pseudo-spectral method, using the truncated
variables ${\bf u}^{>}$ and ${\bf b}^{<}$.  This procedure has to be 
repeated for every value of $K$ for
which the flux needs to be computed. The rest of the fluxes defined 
in Eqs.~(\ref{eq:flux_uinbin})-(\ref{eq:flux_binbout})
 and also the transfer rates defined in 
Eqs.~(\ref{eq:bshell_ushell})-(\ref{eq:bshell_bshell})
are similarly computed. 

In the following section we describe the results of our simulations.

\section{Results}
\label{s:results}

\subsection{Generation of Steady State}
\label{subs:steady_state}

The computational time required to obtain a statistically steady
state on a grid of size $512^{2}$ is large. So to obtain a
steady state in simulations on this grid we proceed
in stages. We run on  a grid of size $64^{2}$ 
till a steady state is achieved. This steady state field is then used
as the initial condition to achieve a steady state in a simulation on a 
grid of size $128^{2}$ and so on to grids $256^{2}$ and then $512^{2}$.

It is theoretically expected that in 2-D, in the long term, the magnetic energy
{\it will} decay \cite{Zeldovich57} even if kinetic energy is stationary.
In our simulations we found that the fields remain steady for the period
over which the simulation was performed (Fig. 1).
 Since the $N=512^{2}$ 
simulations are computationally expensive, we have instead shown the decay of 
magnetic energy for a $N=128^{2}$ simulation. The magnetic energy is found
to have a quasi-steady state over the time interval extending from
approximately $t=20---30$
and it decays over the remaining period, as is theoretically expected.
For the $N=512^{2}$ simulation too, after the period of stationarity, 
the simulations should show
decay of magnetic energy. So, strictly speaking, our steady state is only a
quasi-steady state. In Fig. 1
  we show the magnetic and kinetic energy in 
a quasi-steady state. The Alfv\'{e}n ratio (the ratio of kinetic to magnetic
energy) is found to fluctuate between the values $0.4$ and $0.56$. Hence over the
steady state magnetic energy dominates over the kinetic energy. 
We compute the fluxes and the shell-to-shell transfer
rates over this quasi-steady state once in every unit of non-dimensional 
time. The fluxes
and transfer rates are shown in this paper after averaging  over 15 time 
units in quasi-steady state.

\subsection{Flux studies}
\label{subs:flux_results}

In our numerical simulations we have computed all
the energy fluxes defined in 
Eqs.~(\ref{eq:flux_uinbin})-(\ref{eq:flux_binbout}).
In this we section we discuss these energy fluxes (cascade rates).

In 
Fig. 2
 we show all the fluxes. The total flux $\Pi_{tot}$ 
is positive indicating that there is a net loss of energy from the 
$K$-sphere to 
modes outside for all $K$. In the wavenumber region $25<K<50$, the total 
flux is seen to be approximately constant. This wave number region 
is the inertial range.

The net transfer from kinetic to magnetic energy is a sum of the fluxes 
$\Pi^{u<}_{b<}(K)$, $\Pi^{u<}_{b>}(K)$, $\Pi^{u>}_{b<}[= -\Pi^{b<}_{u>}(K)]$,
and $\Pi^{u>}_{b>}(K)$. In our $512^{2}$ simulations, the sphere of 
radius $K_{max}=241$  encloses all the modes.  We observe from
Fig. 2 that 
the fluxes $\Pi^{u<}_{b>}(K_{max})$, $\Pi^{u>}_{b<}(K_{max})$ and
$\Pi^{u>}_{b>}(K_{max})$ are zero, as there are no modes outside this sphere
of maximum radius $K_{max}$.
The flux $\Pi^{u<}_{b<}(K_{max})$ is found to be positive 
(see Fig. 2),
indicating that there is a net transfer from kinetic energy to magnetic
energy; since the magnetic modes are not being forced, it is this
transfer that keeps the magnetic energy constant in quasi-steady state.  

We now describe features of the fluxes $\Pi^{u<}_{b<}$, 
$\Pi^{u<}_{b>}$, $\Pi^{b<}_{u>}$, $\Pi^{u>}_{b>}$, 
$\Pi^{u<}_{u<}$, and $\Pi^{b<}_{b<}$ observed in our simulations and
plotted in Fig. 2.
We remind the reader that the {\it u}-modes
within a wave number sphere are called the {\it u}-sphere and the
{\it b}-modes within the wave number sphere are called the {\it b}-sphere.
First, we discuss the fluxes that transfer energy between the {\it u}-modes 
and the {\it b}-modes. We find that the flux $\Pi^{u<}_{b<}(K)$ is
positive --- hence, kinetic energy is lost by a {\it u}-sphere to the 
corresponding {\it b}-sphere. The flux $\Pi^{u<}_{b>}(K)$  is also positive. 
It means that a {\it u}-sphere loses energy to the modes outside the
{\it b}-sphere.  We find that the flux $\Pi^{b<}_{u>}(K)$ is negative, which 
implies that the {\it b}-sphere gains energy from modes outside the 
{\it u}-sphere. 
Thus, all these fluxes result in a transfer of 
kinetic energy to magnetic energy. However, $\Pi^{u>}_{b>}(K)$ is
negative,
implying that there is some feedback of energy from modes outside the
{\it b}-sphere to modes outside the {\it u}-sphere. 
The {\it net} transfer, however, is from kinetic to magnetic.  
We shall later show that the 
flux $\Pi^{u>}_{b>}(K)$  plays a crucial role in driving the 
kinetic-to-kinetic flux $\Pi^{u<}_{u>}(K)$.

The energy gained by the {\it b}-spheres is found to be constant for
the approximate range $20 \leq K  \leq K_{max}$. This can be seen from 
the plot of 
$\Pi^{b<}_{u<}(K)+ \Pi^{b<}_{u>}$ (which is the negative of the
total transfer of
kinetic energy to the modes within the {\it b}-sphere). The constancy of
the flux implies that a {\it b}-sphere of radius $K$ and that of radius
of $K+\Delta K$ get the same amount of energy from the {\it u}-modes.
Thus there is no {\it net} energy transfer from the
{\it u}-modes into a {\it b}-shell (of thickness $\Delta K$) beyond
approximately $K=20$.  We therefore conclude that the {\it net} energy transfer
from {\it u}-modes to the {\it b}-sphere occurs within the $K \leq 20$ 
sphere.

We find that there is a kinetic energy gain by the {\it u}-sphere
from {\it u}-modes outside due to the fact that flux $\Pi^{u<}_{u>}(K)$ is
negative --- this is consistent with the numerical simulations
of Ishizawa and Hattori ~\cite{Ishizawa198}. 
This behaviour of kinetic energy is known as an `inverse cascade' in
literature and is reminiscent of the inverse 
cascade of kinetic
energy in 2-D fluid turbulence ~\cite{Lesieur} and of mean square vector 
potential in 2-D MHD turbulence ~\cite{Pouquet78,Biskamp93}.
Note, however that the kinetic energy in 2-D MHD turbulence is not an 
inviscid invariant. Fig. 2
 shows that the inverse cascade of 
kinetic energy exists
in the wave number range $K \le 60$. Even the modes that are being forced
($4 \le k \le 5$) gain energy from higher wave numbers {\it u}-modes. 
The
source of this energy is the flux $\Pi^{b>}_{u>}(K)[=-\Pi^{u>}_{b>}(K)]$ 
which transfers energy from higher {\it b}-modes to the higher {\it u}-modes
and thus  effectively forces them.

We observe from Fig. 2
that there is a loss of magnetic energy from the 
{\it b}-sphere to the {\it b}-modes outside [see $\Pi^{b<}_{b>}(K)$]
,i.e., a forward  cascade of magnetic energy --- this
was also observed in the numerical simulations of Ishizawa and Hattori
 ~\cite{Ishizawa198}.
 Our observation of a forward cascade of magnetic energy contradicts 
the result of the EDQNM closure calculations which predicts an inverse 
cascade of magnetic energy  ~\cite{Pouquet78}.

The flux $\Pi^{b<}_{b>}(K)$ is found to be constant in the 
inertial range.  Using similar reasoning to that given above, we can
conclude that the net magnetic energy transfer to a wave number shell in
the inertial range is zero. However, in this case it would imply that the
{\it entire} energy gained by the shell (of thickness $\Delta K$) from the 
{\it b}-sphere of radius $K$ is lost {\it completely} to 
the modes outside the region $K+\Delta K$.
Thus the magnetic energy cascades down to higher wave numbers 
independently of the transfers between kinetic and magnetic energy in the
inertial range, which are quite significant (see below).
This flux of magnetic energy is sustained by the fluxes 
$\Pi^{u<}_{b<}(K)$ and $\Pi^{u>}_{b<}(K)$, both of  which transfer kinetic
energy into the small {\it b}-spheres.

Figure 3 schematically illustrates the energy fluxes of
Fig. 2 for a $K$ sphere of radius
$K=20$, which is within the inertial range. The energy input due
to forcing and the  small inverse
cascade [$\Pi^{u<}_{u>}(K)$] into the {\it u}-sphere
from higher {\it u}-modes, provides the 
energy input into the {\it u}-sphere. This energy is transferred into and 
outside the {\it b}-sphere by $\Pi^{u<}_{b<}(K)$ and $\Pi^{u<}_{b>}(K)$, 
the latter transfer being the most significant of all transfers 
(see Figs. 2 and 3). 
The energy transferred into the {\it b}-sphere from the {\it u}-sphere 
[$\Pi^{u<}_{b<}(K)$], and a negligible input from the modes outside the 
{\it u}-sphere [$-\Pi^{b<}_{u>}(K)$], cascades down to the higher wave 
number {\it b}-modes. In the higher {\it b}-modes, this cascaded energy 
[$\Pi^{b<}_{b>}(K)$], together with the transfer from the {\it u}-sphere, 
is partly dissipated and partly fed back to the high wave number 
{\it u}-modes. This feedback to the kinetic energy is mostly dissipated, 
though a small inverse cascade takes some energy back into the {\it u}-sphere.
This is the qualitative picture of the energy transfer in 2-D MHD 
turbulence.

The net transfer to each of the four corners of the 
Fig. 3 sum to zero within
the statistical error (which is computed from the standard 
deviation of the sampled data). This is consistent with a quasi-steady-%
state picture.

The results presented here for the quasi-steady-state in a forced
turbulence remain qualitatively valid even for a decaying case --- the
direction of the various fluxes for the decaying case are identical
to that for the forced simulation; but as the energy decays, the magnitudes of 
all the fluxes reduce.

The fluxes give us information about the overall energy transfer from a 
wave number sphere or outside-sphere to another sphere or outside-sphere.
To obtain a more detailed account of the energy transfer, energy exchange
between the wave number shells are now studied. In the
following section we present  a discussion on the 
shell-to-shell energy transfer rates in MHD turbulence.

\subsection{Shell-to-Shell energy transfer-rate studies}
\label{subs:shell_results}

Significant details of energy transfers are revealed by calculating
the shell-to-shell energy transfer rates $T^{uu}_{mn}$, $T^{bb}_{mn}$, 
$T^{bu}_{mn}$ defined in
Eqs.~(\ref{eq:ushell_ushell})-(\ref{eq:bshell_ushell}).
We partition the k-space into shells at wave numbers $k_{n} (n=1,2,3,...)
=1,16,19.02,22.62,...,2^{(n+14)/4}$.
The first shell extends from $k_{1}=1$ to
$k_{2}=16$ --- a division of the wave number space into smaller shells at 
the these lower wavenumbers
 will contain too few modes; the second shell extends
from $k_{2}=16$ to $k_{3}=19.02$, ..., the $m^{th}$ shell extends from
$k_{m}$ to $k_{(m+1)}$.
Thus, the effective shell-to-shell energy transfer rate  from the 
$n^{th}$ {\it u}-shell 
to the $m^{th}$ {\it u}-shell [Eq.~(\ref{eq:ushell_ushell})] 
can be written as,
\begin{equation}
T^{uu}_{mn} = \sum_{k_{m}<k<k_{m+1}} \sum_{k_{n}<p<k_{n+1}} 
               \sum_{\bf q}^{\Delta} {\cal{\slash{S}}}^{uu}({\bf k|p|q}) ,
\label{eq:shell_utou}
\end{equation}
and the effective shell-to-shell energy transfer rate  from the 
$n^{th}$ {\it b}-shell to the $m^{th}$ {\it b}-shell
[Eq.~(\ref{eq:bshell_bshell})] can be written as 
\begin{equation}
T^{bb}_{mn} = \sum_{k_{m}<k<k_{m+1}} \sum_{k_{n}<p<k_{n+1}} 
               \sum_{\bf q}^{\Delta} {\cal{\slash{S}}}^{bb}({\bf k|p|q}) .
\label{eq:shell_btob}
\end{equation}
and the effective shell-to-shell energy transfer rate from 
the $n^{th}$ {\it u}-shell to the $m^{th}$ {\it b}-shell, as defined
in Eq.~(\ref{eq:bshell_ushell}) can be written as, 
\begin{equation}
T^{bu}_{mn} = \sum_{k_{m}<k<k_{m+1}} \sum_{k_{n}<p<k_{n+1}} 
               \sum_{\bf q}^{\Delta} {\cal{\slash{S}}}^{bu}({\bf k|p|q}) .
\label{eq:shell_btou}
\end{equation}

We categorise the energy transfer between {\it u}-shells and
the {\it b}-shells as being {\it homologous} if the transfers are
between the corresponding shells (of the same wave number range 
$k_{n}<k<k_{n+1}$, say). Transfers between different shells are therefore
{\it non-homologous}. Further, transfers (kinetic-to-magnetic, 
kinetic-to-kinetic,
and magnetic-to-magnetic) involving shells which are close in wave number
space are called {\it local} transfers, as is the convention. Two
shells are considered close if the wave number ratio of the larger 
shell to the smaller shell is less than 2 (in this study this would
include the shells between $n+4$ to $n-4$ from the $n^{th}$ shell). Transfers
involving shells more distant are called {\it non-local}.

In Fig. 4
we plot the energy transfer rates $T^{bu}_{mn}$ between 
{\it u}-shells and  {\it b}-shells. 
It is evident from the figure that 
the transfer rates between shells in the inertial range are virtually
independent of the individual values of the indices
$m$ and $n$, and only dependent on their differences. This means that the
transfer rates in the inertial range are {\it self-similar}. The differences
in $T^{bu}_{mn}$ for various $n$ are smaller than the standard deviation
of the sampled data, indicating that the perceived self-similarity is 
statistically significant.

We now discuss our simulation results for the non-homologous transfer
rates between {\it u}-shells and {\it b}-shells. 
In Fig. 4 we have
shown these transfer rates from the $n^{th}$ {\it u}-shell to the $m^{th}$
{\it b}-shell by plotting $T^{bu}_{mn}$ versus $m$ for various values of $n$.  
We find that for all $m$, except for $m=n-1$ and $n$, $T^{bu}_{mn}$ is 
positive. This implies that a {\it u}-shell loses energy to  
all the {\it b}-shells but gains energy from the 
$(n-1)^{th}$ and $n^{th}$ {\it b}-shells. The quantity
$T^{bu}_{mn}$ is found to be small for $m < n$ 
in comparison with the transfer for $m > n$. Consequently, energy
from a {\it u}-shell is mainly transferred to the {\it b}-shells at 
higher wave numbers. We had found earlier that the flux
$\Pi^{u>}_{b<}(K)$ is much smaller in magnitude to the flux $\Pi^{u<}_{b>}(K)$.
The fluxes and the transfer rates are consistent with each other.
From Fig. 4
we also find that the energy transfer rate $T^{bu}_{1n}$ from the $n^{th}$
{\it u}-shell to the $1^{st}$ {\it b}-shell is positive, implying that
the $1^{st}$ {\it b}-shell gains energy from the {\it u}-shells ---
this is a nonlocal transfer of energy from the large {\it u}-modes
to the small {\it b}-modes. The flux
$\Pi^{b<}_{u>}$ observed in Section~\ref{subs:flux_results} 
(see Fig. 2)
is due to this nonlocal transfer.

The homologous transfers are found to transfer energy from {\it b}-shells to 
{\it u}-shells (see Fig.4); this is in contrast to 
the net energy transfer
and a majority of the non-homologous transfers which are from kinetic
to magnetic.
We find  that the energy gained by  a 
{\it u}-shell  through the
homologous transfers is larger than the {\it total} loss of energy by
the {\it u}-shell through
non-homologous transfers. Consequently, there is a {\it net} gain of energy
by the  {\it u}-shells in the inertial range. In
Section~\ref{subs:flux_results} it was
shown that the magnetic energy outside a {\it b}-sphere is lost to 
kinetic energy outside the {\it u}-sphere. 
It is now clear that this transfer arises primarily due to the homologous
transfers from magnetic to kinetic. 

The Fig. 4 shows energy transfer rates
$T^{bu}_{mn}$, to {\it b}-shells from {\it u}-shells indexed as $n=5,6,7,8,9$.
In Fig. 5
we show the transfer rates $T^{bu}_{mn}$ from the $n=1$ 
{\it u}-shell to all the {\it b}-shells. Recall that the first {\it u}-shell 
comprises the small wave number modes, $k=1$ to $16$.
Comparing the magnitudes of $T^{bu}_{mn}$ to the {\it b}-shells from
the first {\it u}-shell (see Fig. 5)
and from the {\it u}-shells at higher wavenumbers 
(see Fig. 4), we see that the energy transfer rate
from the $1^{st}$ {\it u}-shell dominates the transfers from all other
{\it u}-shells. Hence, there is a large amount of
{\it non-local} transfer from the first {\it u}-shell to the {\it b}-shells.
In Section~\ref{subs:flux_results} we had claimed that there is no net
kinetic energy transfer into any inertial range {\it b}-shell. 
Now we show this explicitly in Fig. 6
by plotting the net kinetic energy transferred into a {\it b}-shell 
(=$ \sum_{n} T^{bu}_{mn}$) versus the shell index $m$ ($\diamond$ in the 
figure).  From the figure, the net energy transferred into the 
inertial range {\it b}-shell can be seen to be nearly zero. 
We also plot in Fig. 6
the net kinetic energy transfer into a {\it b}-shell from all 
{\it u}-shells, {\it except} the first ($+$ in the figure). This
quantity is now no longer zero and has a significant magnitude.
Therefore, the non-local transfer from the first shell plays an important 
part in balancing the other kinetic energy transfers into an inertial range 
{\it b}-shell.

We have also computed the transfer rates of magnetic energy $T^{bb}_{mn}$
from $n^{th}$ to $m^{th}$ {\it b}-shell. In Fig. 7
we have plotted the quantity $T^{bb}_{mn}$ versus $m$ for different 
values of $n$ in the inertial range. 
We find that the differences in $T^{bb}_{(n+\Delta n)n}$ 
for any fixed $\Delta n$(=$n-m$) are smaller than the standard deviation of 
their statistical fluctuations computed from the sampled data. 
Hence we can conclude that $T^{bb}_{mn}$ is self-similar in the
inertial range, 
dependent only on the difference $\Delta n$ and independent of the location 
of the shell $n$. We find that the transfer rates $T^{bb}_{mn}$ are 
negative for $m < n$ and they are positive for $m > n$ 
(see Fig. 7). 
Hence a {\it b}-shell gains energy from the
{\it b}-shells of smaller wave numbers and loses energy to the {\it b}-shells 
of larger wave numbers.
Since $T^{bb}_{mn}$ is self-similar, the energy lost from a shell
($n-\Delta n$) to $n$ is equal to the energy lost from $n$ to the shell 
($n+\Delta n$). Thus, the net magnetic energy transfer into any inertial
range shell is zero, and the energy cascades from the smaller 
wave numbers to the higher wave numbers.

We now discuss the simulation results for the kinetic energy transfer rates
$T^{uu}_{mn}$ from the $n^{th}$ to the $m^{th}$ shell. The results are shown
in Fig. 8 where we have plotted $T^{uu}_{mn}$ 
versus $m$ for various values of $n$. We find that the most dominant 
transfers are from the $n^{th}$ {\it u}-shell to $m=n \pm 1$. From the sign 
of these transfers we see 
that kinetic energy is gained from $(n-1)^{th}$ shell and 
lost to $(n+1)^{th}$ shell. 
This means that the {\it local} transfers from the adjacent shells
result in a {\it forward} cascade of energy towards the {\it large} wave 
numbers. The transfers from other shells are largely negligible except
for the transfer to the shell $m=1$ (shown boxed on the left of the figure),
which represents a loss from high wave number modes to the $m=1$ shell.
This {\it non-local} transfer (to the first shell)  produces an
{\it inverse} cascade to the {\it small} wave numbers, observed earlier 
in Section~\ref{subs:flux_results}.
Now it is clear that 
this inverse cascade is due to non-local transfers. 
The local and non-local transfers are seen to possess 
altogether different features: the former is a forward cascade which
seems largely self-similar, while the latter is an inverse cascade
{\it only} to the first shell.

We schematically illustrate in Fig. 9
the energy transfer between shells.
In this figure we show directions of the most significant
transfers in the inertial range. The arrows indicate the directions of the
transfers, and thickness of the arrows indicates the approximate 
relative magnitudes. Since the local transfer rates are 
self-similar, the transfers from any other shells in the inertial range 
will also show
the same pattern. An inertial range {\it b}-shell gains significant 
amount of energy from 
the smaller {\it u}-shells through both local and non-local 
transfers, and it also locally gains energy from the smaller 
{\it b}-shells ($T^{bb}_{mn}$). 
The energy  gained by a {\it b}-shell from the smaller 
{\it b}-shells is {\it exclusively}
lost to the larger {\it b}-shells and the energy gained 
from the {\it u}-shells is mainly lost through homologous transfer 
($T^{bu}_{nn}$) to the corresponding {\it u}-shell.
A small fraction of the energy is also lost to the larger {\it u}-shells.
In addition to the energy from the {\it b}-shells, a {\it u}-shell also 
gains energy from smaller {\it u}-shells by {\it local} transfer.
The energy of {\it u}-shells is mainly lost locally to higher {\it b}-shells
and {\it u}-shells, but a significant amount is also transferred to  smaller 
{\it u}-modes ($T^{uu}_{mn}$) by a {\it non-local} inverse cascade.

As illustrated in Fig. 9, there is a transfer 
of energy from the first
{\it u}-shell to the first {\it b}-shell. This is the most significant
gain of magnetic energy from the kinetic energy. Under steady state the
magnetic energy gained by the first shell gets transferred to the higher
{\it b}-shells. This transfer yields the forward cascade of magnetic energy
discussed in Section~\ref{subs:flux_results}

To summarise, we find that there are various types of energy transfers in
MHD turbulence: local, nonlocal, homologous, and non-homologous. The main
local transfers are the forward magnetic energy transfer and the forward
kinetic energy transfer, the homologous transfer from the magnetic to 
the kinetic energy (all in the inertial range). There is nonlocal transfer
from inertial range {\it u}-shells to the first {\it b}-shell and the
first {\it u}-shell, and from the first {\it u}-shell
to the {\it b}-shells. The most significant energy transfer is
the transfer of kinetic energy from the first {\it u}-shell
 to magnetic energy in the first {\it b}-shell.

In this section we had extensively described various cascade rates (fluxes)
and shell-to-shell energy transfer rates. It is clear that the complete
picture is quite complex. In fact, some of the features we have seen 
contradicts earlier conjectures and results.
These differences are discussed in the next section.

\section{Discussion}
\label{s:discussion}
We have investigated the features of kinetic and 
magnetic energy transfer between scales at low values of cross-helicity
in a quasi-steady state of forced 2-D MHD turbulence. Several interesting
observations were made in our simulations.  A summary of the 
results is given below. For the following discussion refer to 
Figs. 3 and 9.

1)  There is a net transfer of energy from the kinetic to the magnetic. 

2) There is an energy transfer to the large-scale magnetic field from
the large-scale velocity field ({\it u}-sphere to {\it b}-shere flux
$\Pi^{u<}_{b<}$ in Fig. 2),
and also from the small-scale velocity field 
(flux from {\it b}-sphere to modes outside the {\it u}-sphere 
$\Pi^{b<}_{u>}$ in Fig. 2). The former transfer
is of a greater magnitude than the latter. 
The magnetic field enhancement is primarily caused
by a transfer from the {\it u}-sphere to the {\it b}-sphere. 
Indeed the first few {\it b}-modes get most of this energy.

3) Other significant transfers, not noted in earlier work, are from
the large-scale velocity field to the small-scale magnetic field
({\it u}-sphere to outside the {\it b}-sphere) and
an interesting reverse transfer from the small-scale magnetic field to the 
small-scale velocity field (from modes
outside the {\it b}-sphere to modes outside {\it u}-sphere). 

4)  There is an {\it inverse} cascade in the velocity field --- this is 
consistent with the observation of Ishizawa and Hattori arising from
numerical simulations ~\cite{Ishizawa198} and is also consistent with
the EDQNM closure calculations ~\cite{Pouquet78,Ishizawa298}. 
This inverse cascade of kinetic energy is driven by 
the reverse transfer of energy from magnetic to the velocity field at the
small scales. Although the flux study points to an inverse cascade of
kinetic energy, the shell-to-shell energy transfer rates reveal the following
feature. There exists both an inverse and a
forward transfer of kinetic energy. The inverse transfer is primarily
{\it nonlocal}, coming from the larger wave numbers to the first few
modes. The forward cascade is {\it local}, i.e., between same sized
eddies and is self-similar in the inertial range. 

5) There is a {\it forward} cascade of magnetic energy towards 
the small scales. This is consistent with other recent numerical
simulations ~\cite{Ishizawa198}. EDQNM closure calculations also yield
a magnetic energy transfer to the small-scales ~\cite{Ishizawa298}. 
The magnetic energy transfer is primarily {\it local} and
is found to be self-similar in the inertial range. 

6) In the inertial range we found a dichotomy between the 
homologous transfers (defined as the transfers between kinetic and magnetic 
energy shells of same wave numbers) and non-homologous transfers (which are 
the transfers between shells of different wave numbers). These transfers
are more complex than had been anticipated ~\cite{Pouquet78} and are summarised
and discussed in detail in Section~\ref{subs:shell_results} 
(also see Fig. 3).

The most important
outcome of our study can be stated in the following points : (1) 
Enhancement of magnetic energy occurs due to the energy transfer
to the large-scale magnetic field from the large-scale velocity field,
(2) there is a forward cascade of magnetic energy from the 
large-scale magnetic field to the small-scale magnetic field.
Our results are important since they clarify some of the proposals made
earlier regarding the physical mechanism for the
generation of large-scale magnetic field. Pouquet ~\cite{Pouquet78}
had found that
the large-scale magnetic energy is destablised by the small-scale
magnetic energy. Ishizawa and Hattori ~\cite{Ishizawa298} proposed that
the large-scale magnetic energy is enhanced due to energy transfer
from the small-scale velocity field to the large-scale magnetic
field. Although we find such a transfer, the dominant transfer to the
large-scale magnetic energy is from the large-scale kinetic energy.
Pouquet and Patterson ~\cite{PouquetPatt78}
proposed in context of 3-D turbulence that there an inverse 
cascade of energy from small-scale to large-scale magnetic field.
However, our numerical calculations give a forward cascade in 2-D turbulence.
A simulation should be done to verify whether this feature is
also present in 3-D turbulence.

The EDQNM closure calculations yield a net transfer of energy to
the large-scale magnetic field from the small-scales (magnetic+kinetic)
 ~\cite{Pouquet78,Ishizawa298}.
In our simulations, the energy lost by the large-scale magnetic
field to the small-scale magnetic field is much larger than the
energy gained by the large-scale magnetic field from the small-scale
velocity field. Hence, contrary to the predictions of EDQNM calculations,
the large-scale magnetic field loses energy to the small-scales. A
similar observation was also made by Ishizawa and Hattori in their
simulations ~\cite{Ishizawa198}. Thus, it appears that EDQNM calculations
do not yield the correct strengths of the various transfers.

  The picture obtained by us for the energy fluxes and the
shell-to-shell energy transfers are consistent and complement each other.
The fluxes and the transfer rates discussed here could find 
applications in the dynamo problem. In astrophysical objects, like
galaxies and the sun, the magnetic field is thought to have arisen due to the
amplification of a seed magnetic field. Our study has been performed over a
quasi-steady state with a low Alfv\'{e}n ratio. In order to understand the
build-up of magnetic energy starting from a seed value it is important to
perform a similar study at high Afv\'{e}n ratio.
In some of the popular models, like the $\alpha$-dynamo \cite{Moffatt},
the mean magnetic field gets amplified in presence of helical
fluctuations. 
It must be borne in mind that
our caclulations are two-dimsensional and devoid of magnetic helicity
and kinetic helicity.  A three-dimensional calculation
(with the inclusion of magnetic and kinetic helicities) of
various fluxes and shell-to-shell energy transfer rates will yield
important insights on some unresolved issues concerning
enhancement of magnetic energy.

The fluxes also find important applications in various phenomenological
studies. For example, Verma {\it et al.} \cite{MKV95a} estimated the
turbulent dissipation rates in the solar wind and obtained the
temperature variation of the solar wind as a function of solar
distance. The various cascade rates discussed here could be useful
for various astrophysical studies. For example, $\Pi^{u<}_{b<}$ and
$\Pi^{u<}_{b>}$ can be used for studying the variation of $r_{A}$
of the solar wind.

The physics of MHD turbulence is still unclear. 
The studies of various
fluxes and transfer rates shed light at various aspects which will
help us in getting a better understanding of MHD turbulence.

\acknowledgments

We thank Prof. R. K. Ghosh of Computer Science Dept., Indian Institute of
Technology (IIT) Kanpur,
 for providing us computer time through the project
TAPTEC/COMPUTER/504 sponsored by All India Council for Technical 
Education (AICTE). 


\bibliography{ref}

\newpage

\begin{center}
FIGURE CAPTIONS
\end{center}

\vspace{2.0cm}

\noindent Fig. 1
Evolution of the total kinetic energy and the 
magnetic energy for simulations on grid sizes $512^{2}$ and $128^{2}$.
For $512^{2}$ a quasi-steady magnetic energy is obtained over the
period of the simulation. It is demonstrated in the $128^{2}$ simulation
that a quasi-steady magnetic energy eventually decays --- a quasi-steady
magnetic energy is obtained from $t=60$ to $100$.

\vspace{1.0cm}

\noindent Fig. 2 
The simulation results of fluxes. The following fluxes have
been plotted :
the total flux ($\Pi_{tot}$), 
the kinetic energy flux from {\it u}-sphere to outside {\it u}-sphere 
$\Pi^{u<}_{u>}$,
the magnetic energy flux from {\it b}-sphere to outside {\it b}-sphere 
$\Pi^{b<}_{b>}$,
the energy flux from {\it u}-sphere to {\it b}-sphere $\Pi^{u<}_{b<}$,
the energy flux from {\it u}-sphere to modes outside the 
{\it b}-sphere $\Pi^{u<}_{b>}$,
the energy flux from {\it b}-sphere to modes outside the 
{\it u}-sphere $\Pi^{b<}_{u>}$,
the energy flux from modes outside {\it u}-sphere to modes outside the 
{\it b}-sphere $\Pi^{u>}_{b>}$,
and the net flux out of a {\it b}-sphere 
($\Pi^{b<}_{u<}+\Pi^{b<}_{u>}$) have been plotted in this figure.

\vspace{1.0cm}

\noindent Fig. 3
The schematic illustration of the 
directions and the magnitudes of the fluxes 
plotted in Fig.~\ref{fig:flux_results} (also see Fig. 3.13
of Chapter 3). Also shown are the magnitudes 
of the kinetic energy input rate due to forcing, and the total 
dissipation of kinetic and magnetic energy.  The fluxes are shown 
for $K=20$ but are representative of the entire inertial range. All
quantities have been time-averaged. The fluctuations of 
the fluxes (except $\Pi^{u<}_{b<}$) 
and the dissipation rate are approximately equal to 0.005. The fluctuations
in $\Pi^{u<}_{b<}$ are higher and is approximately 0.01 .

\vspace{1.0cm}

\noindent Fig. 4
 The energy transfer rate $T^{bu}_{mn}$ from the $n^{th}$ 
 {\it u}-shell to the $m^{th}$ {\it b}-shell.
 The loss of energy from the $n^{th}$ {\it u}-shell 
 to the $m^{th}$ {\it b}-shell is defined to be positive.

\vspace{1.0cm}

\noindent Fig. 5
The energy transfer rate from the $1^{st}$ {\it u}-shell to 
 the {\it b}-shells.

\vspace{1.0cm}

\noindent Fig. 6
The diamonds ($\diamond$) represent the net energy transfer into a 
{\it b}-shell from all the {\it u}-shells. The pluses ($+$) represent
the net energy transfer into a {\it b}-shell from all {\it u}-shells
except the $1^{st}$ one.

\vspace{1.0cm}

\noindent Fig. 7
The energy transfer rate $T^{bb}_{mn}$ from the $n^{th}$ 
{\it b}-shell to the $m^{th}$ {\it b}-shell.  The loss of energy 
from the $n^{th}$ {\it b}-shell to the $m^{th}$ {\it b}-shell is 
defined to be positive.

\vspace{1.0cm}

\noindent Fig. 8
The energy transfer rate $T^{uu}_{mn}$ from the $n^{th}$ 
{\it u}-shell to the $m^{th}$ {\it u}-shell.
The loss of energy from the $n^{th}$ 
{\it b}-shell to the $m^{th}$ {\it b}-shell is defined to be 
 positive. The boxed points represent energy transfer from the 
 $n^{th}$ {\it u}-shell to the $1^{st}$ {\it u}-shell.

\vspace{1.0cm}

\noindent Fig. 9
A schematic representation of the direction and the 
magnitude of energy transfer between {\it u}-shells and 
{\it b}-shells. The relative magnitudes of the different transfers
has been represented by the thickness of the arrows.  The non-local 
transfers with the $1^{st}$ shell have been shown by dashed lines.

\newpage

\pagebreak
\begin{figure}[h]
\vskip -20cm
\psfig{file=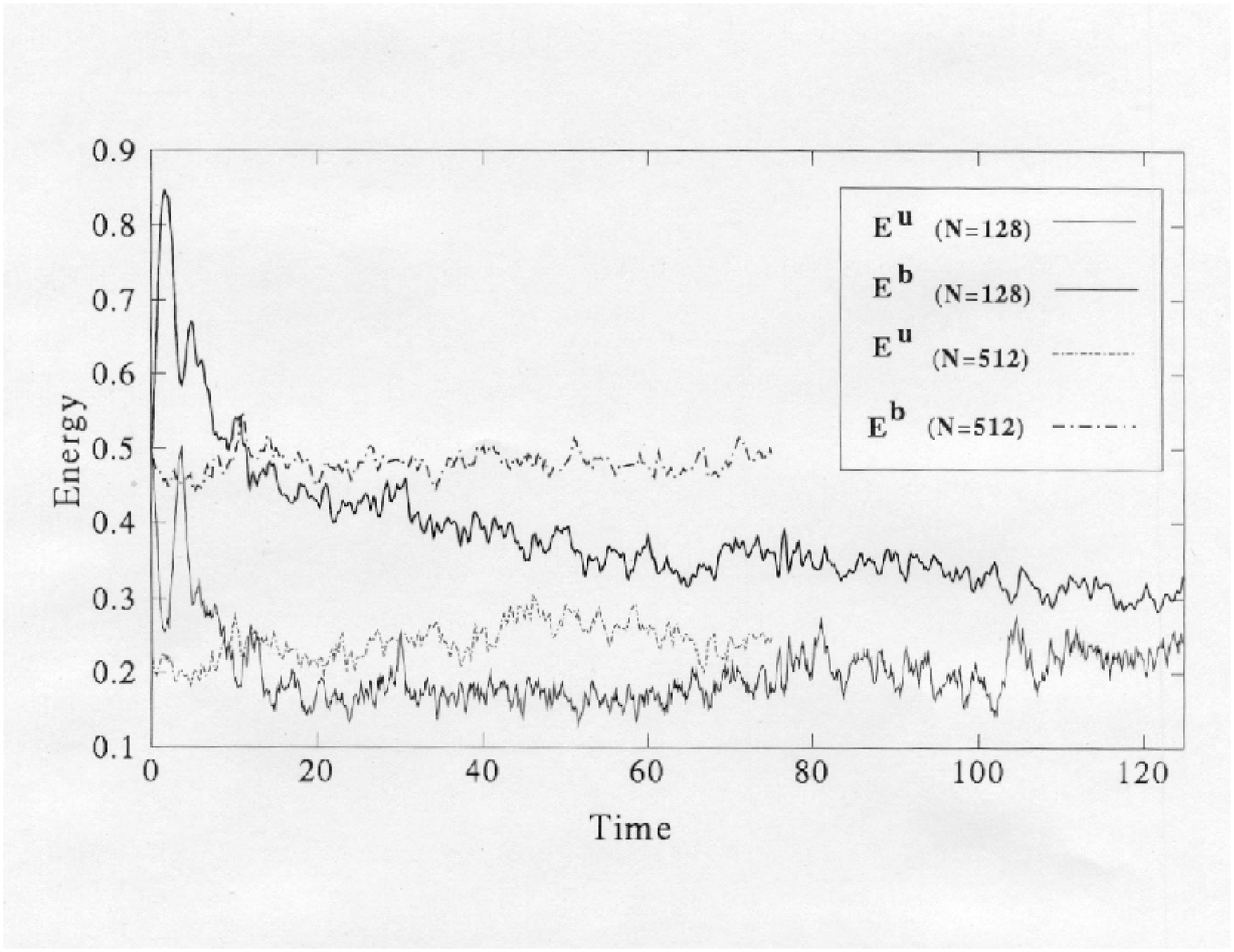,width=0.8\textwidth}
\label{fig:energy}
\caption{}
\end{figure}

\newpage

\begin{figure}[h]
\centerline{\mbox{\psfig{file=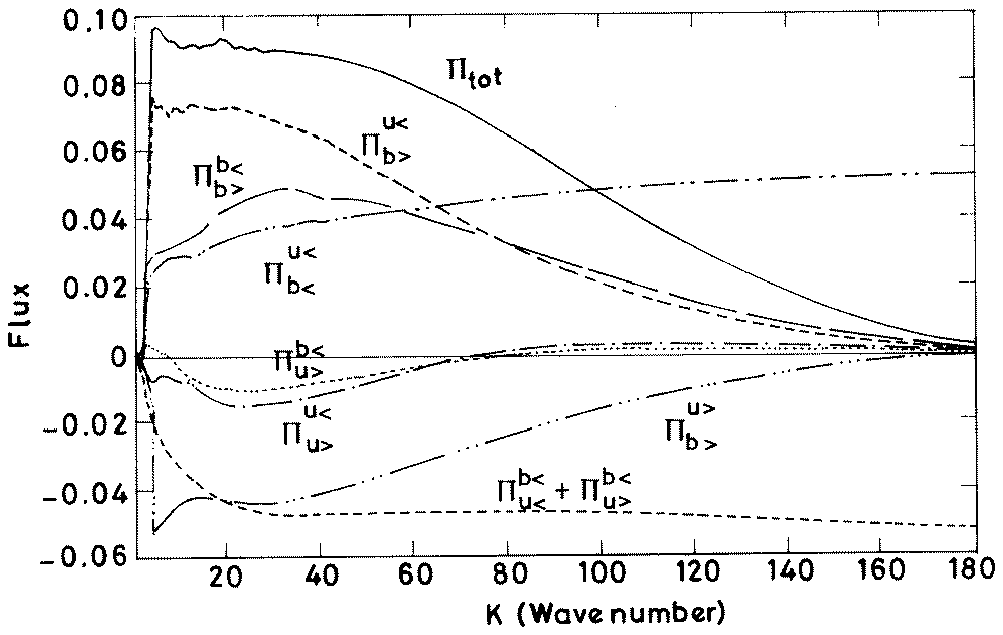,width=0.8\textwidth}}}
\label{fig:flux_results}
\caption{}
\end{figure}

\begin{figure}[h!]
\centerline{\mbox{\psfig{file=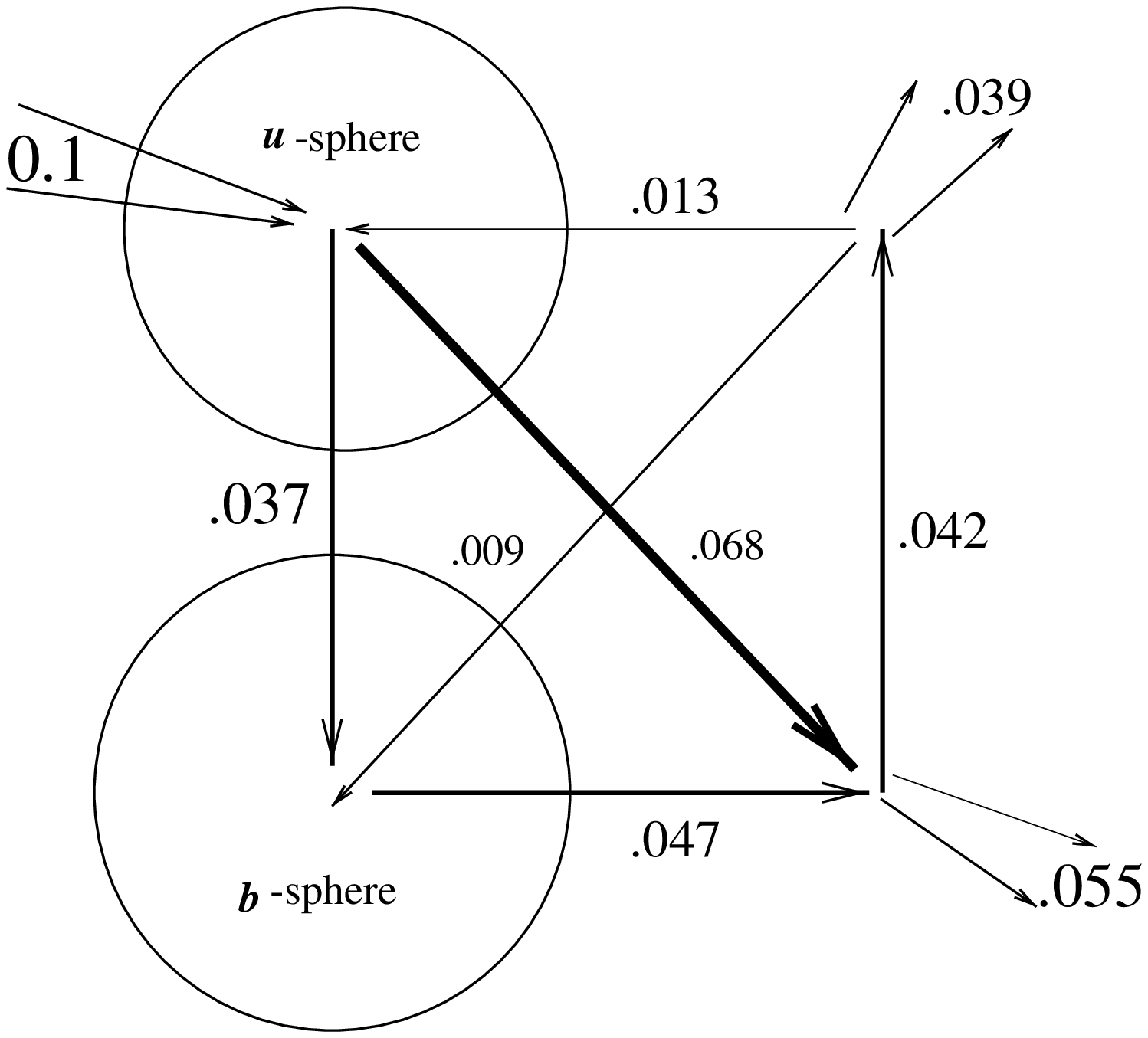,width=0.8\textwidth}}}
\label{fig:flux_schem}
\caption{}
\end{figure}

\begin{figure}[h]
\centerline{\mbox{\psfig{file=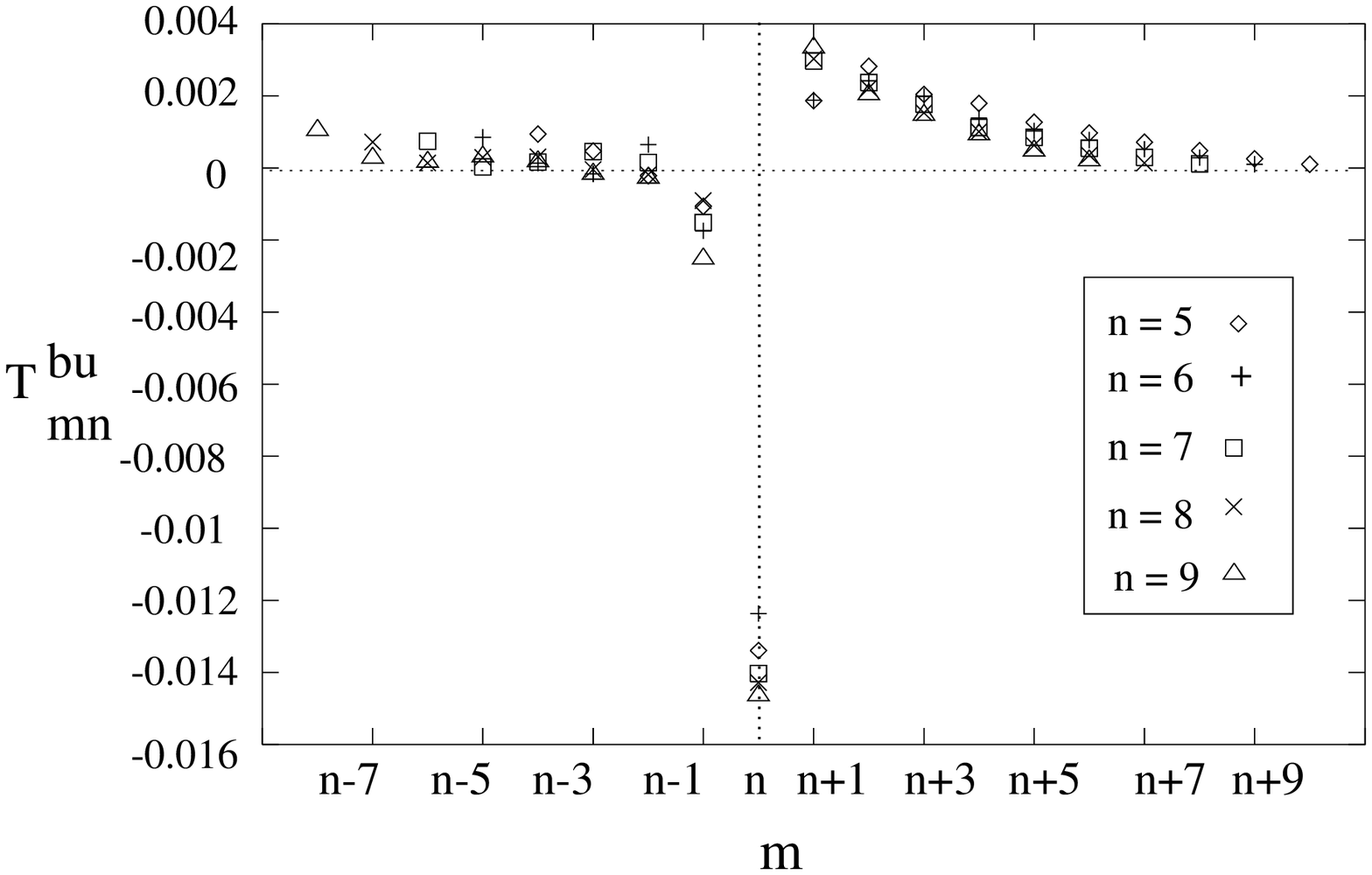,width=0.8\textwidth}}}
\label{fig:ushell_bshell}
\caption{}
\end{figure}

\begin{figure}[h]
\centerline{\mbox{\psfig{file=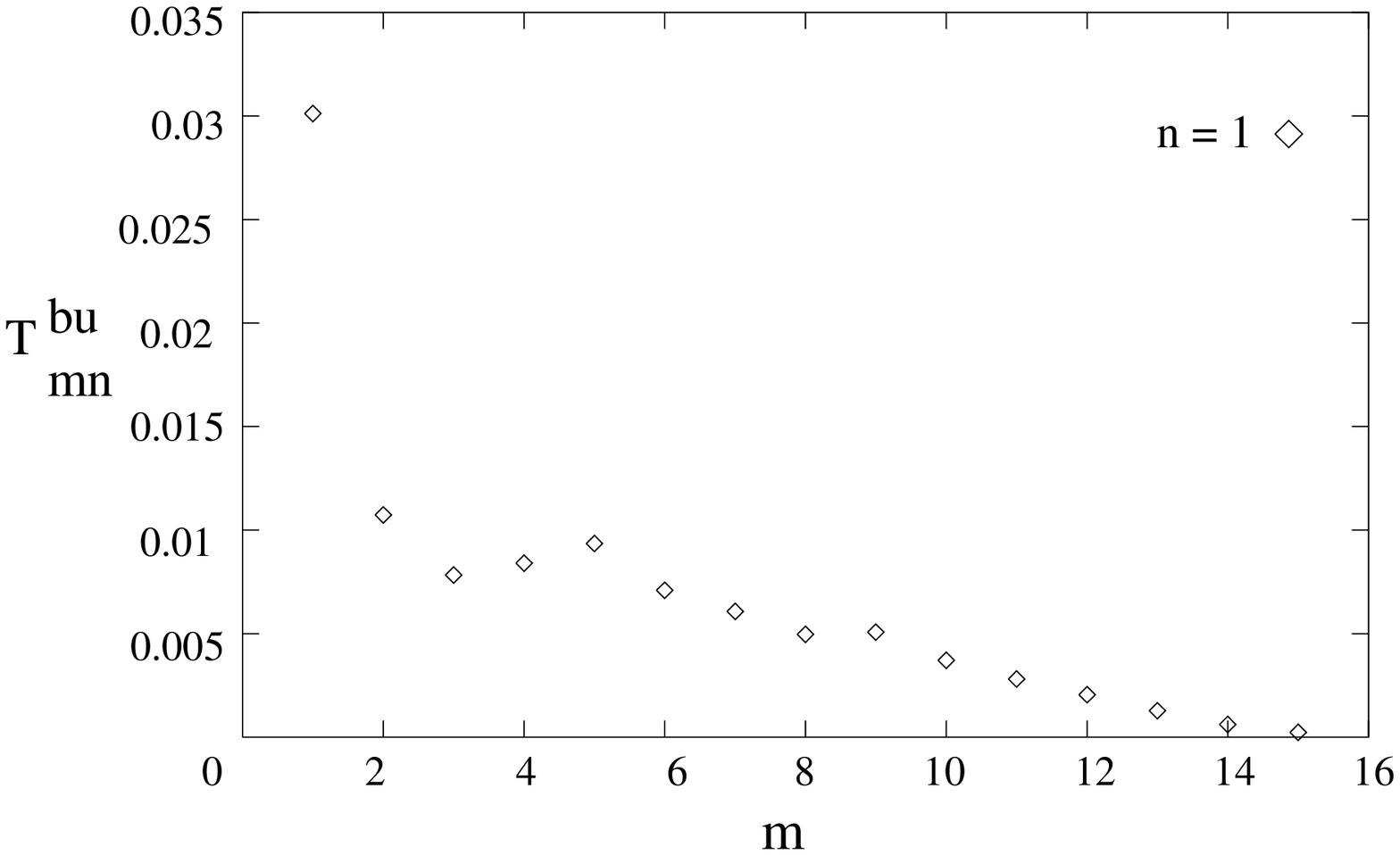,width=0.8\textwidth}}}
\label{fig:u1_to_allb}
\caption{}
\end{figure}

\begin{figure}[h]
\centerline{\mbox{\psfig{file=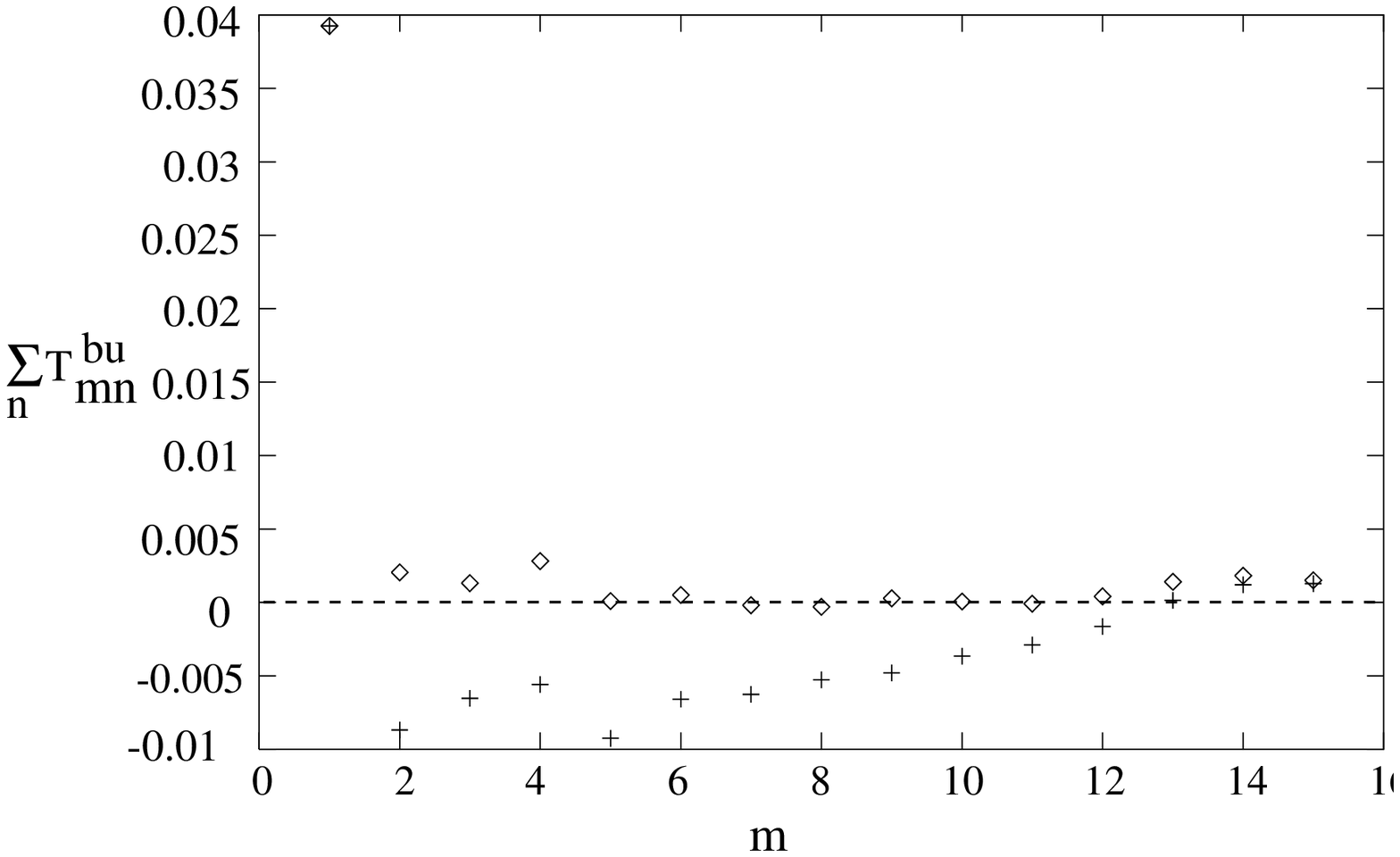,width=0.8\textwidth}}}
\label{fig:uall_to_binertial}
\caption{}
\end{figure}

\begin{figure}[h]
\centerline{\mbox{\psfig{file=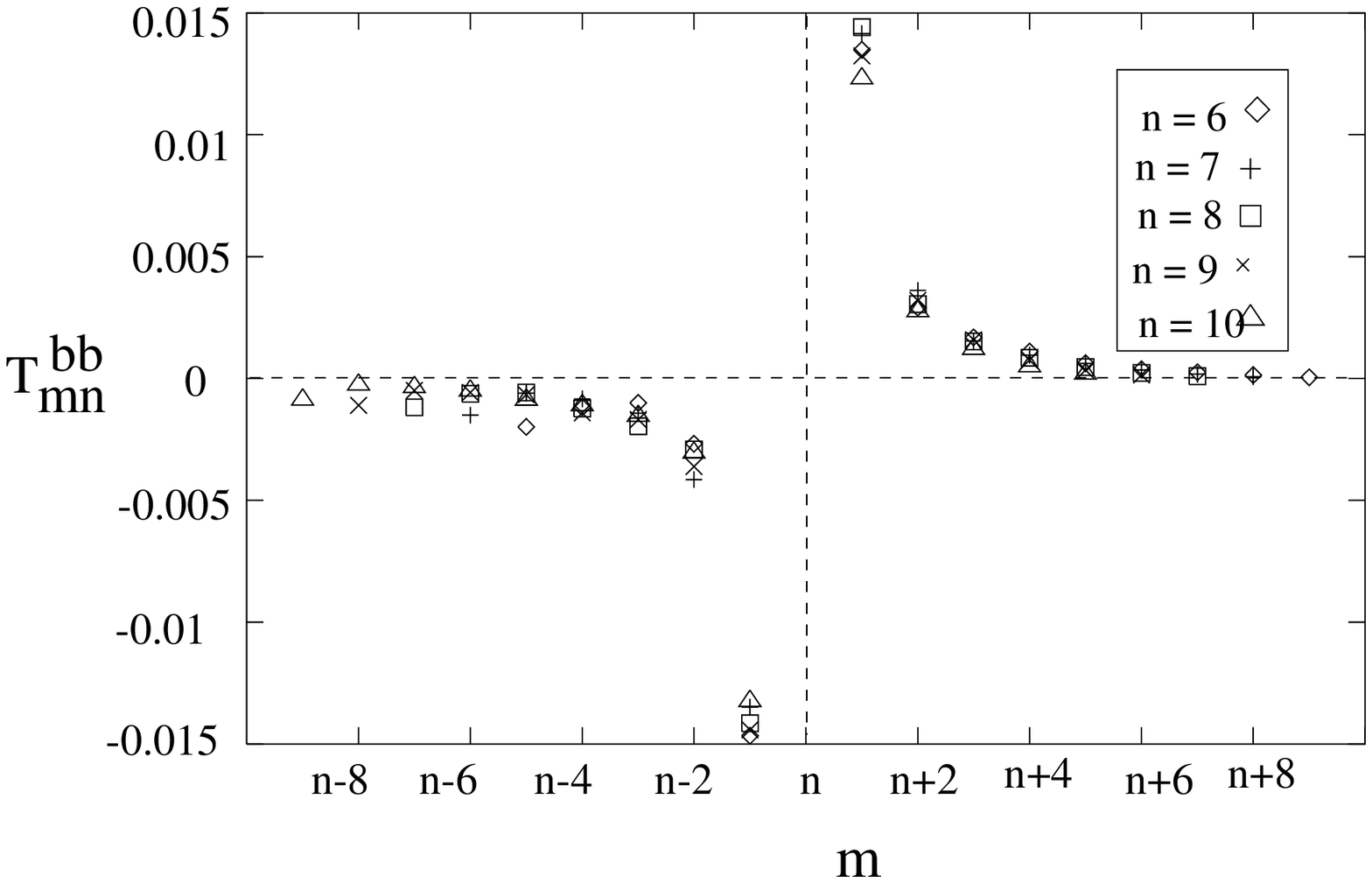,width=0.8\textwidth}}}
\label{fig:bshell_bshell}
\caption{}
\end{figure}

\begin{figure}[h]
\vspace{-2.0cm}
\centerline{\mbox{\psfig{file=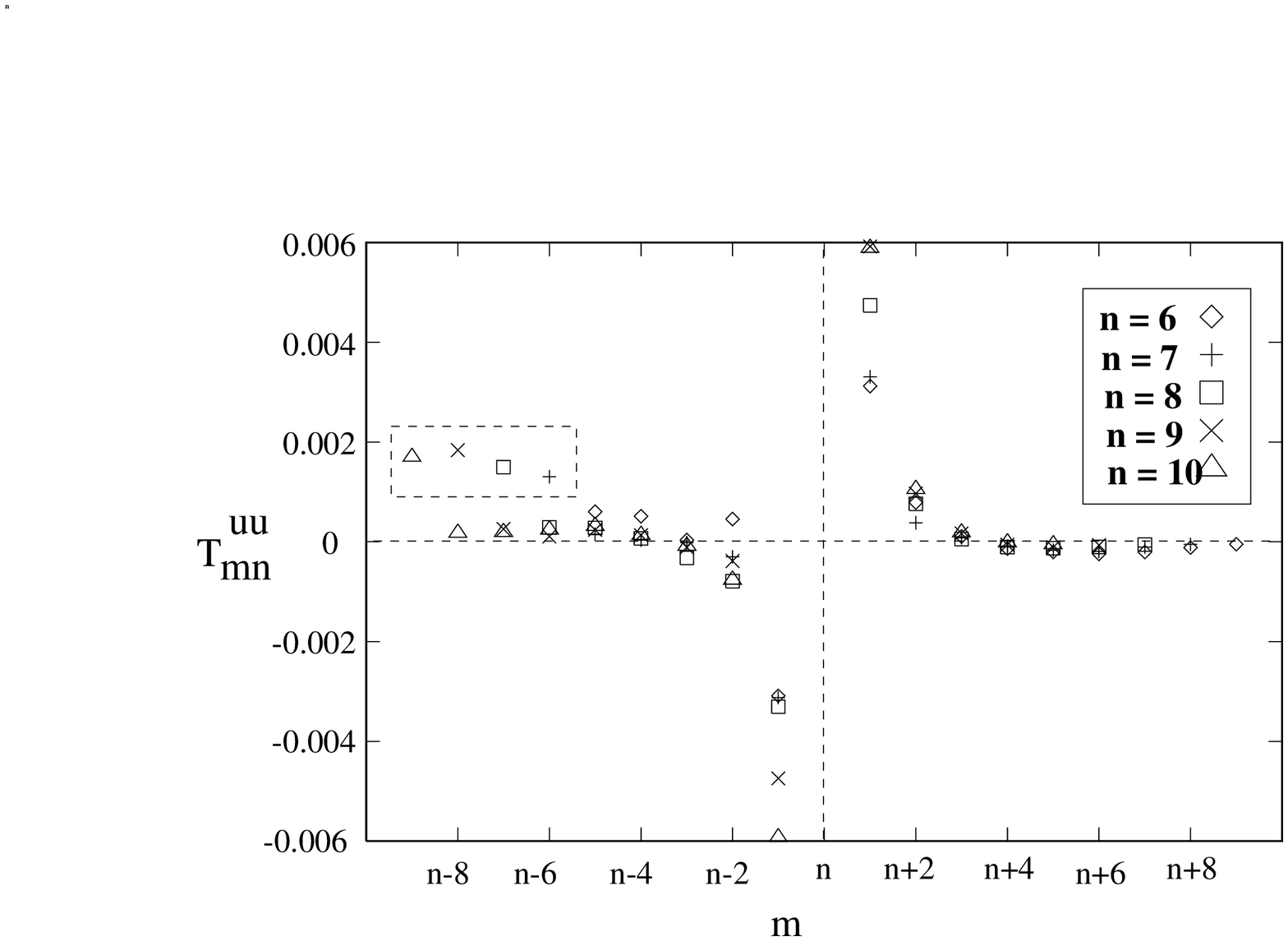,width=0.8\textwidth}}}
\label{fig:ushell_ushell}
\caption{}
\end{figure}

\begin{figure}[h]
\centerline{\mbox{\psfig{file=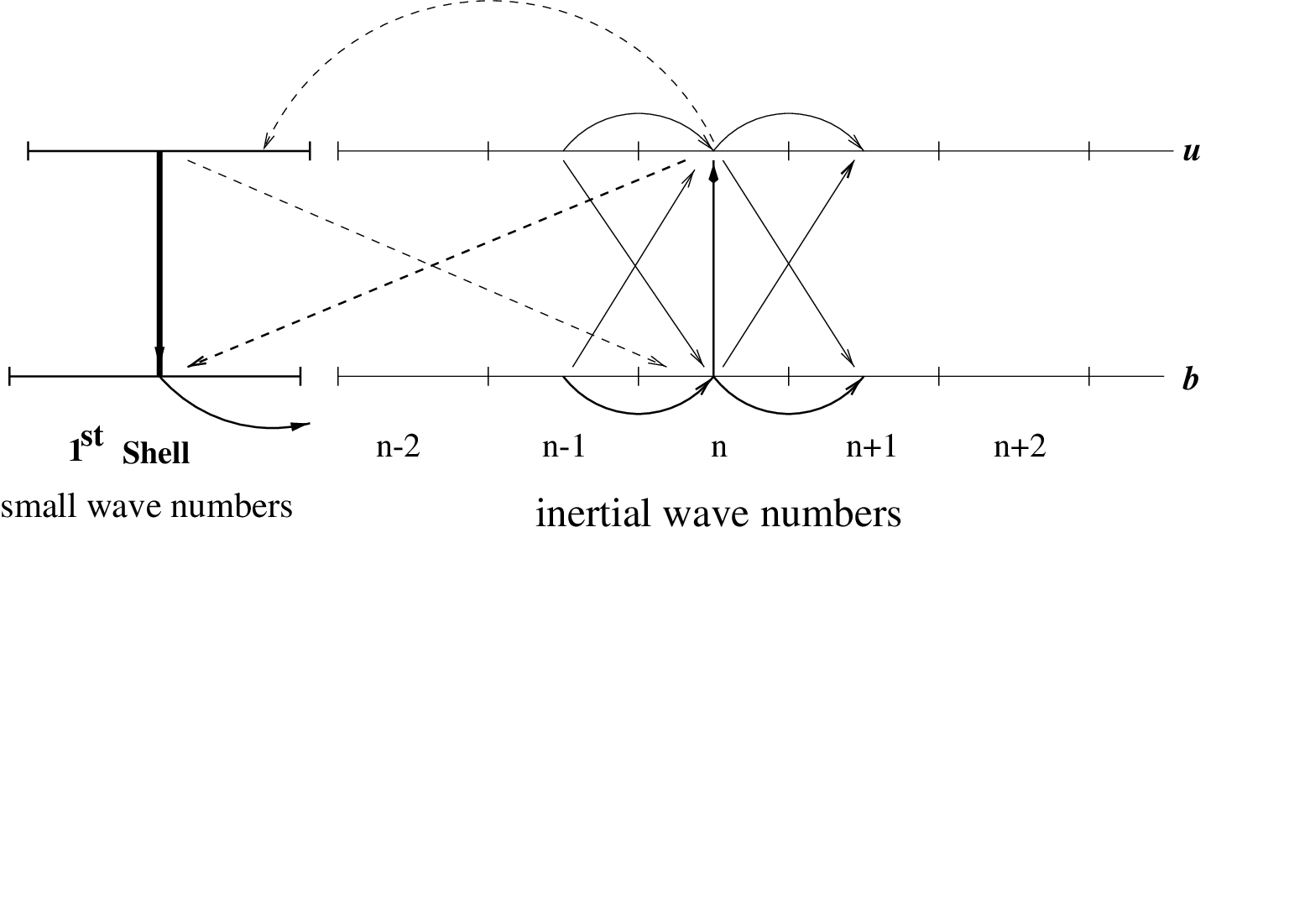,width=0.9\textwidth}}}
\label{fig:shell_schem}
\caption{}
\end{figure}

\end{document}